\documentclass[graybox]{svmult}

\usepackage{todonotes}
\usepackage{mathptmx}       % selects Times Roman as basic font
\usepackage{helvet}         % selects Helvetica as sans-serif font
\usepackage{courier}        % selects Courier as typewriter font
\usepackage{type1cm}        % activate if the above 3 fonts are
\usepackage{makeidx}         % allows index generation
\usepackage{graphicx}        % standard LaTeX graphics tool
\usepackage{multicol}        % used for the two-column index
\usepackage[bottom]{footmisc}% places footnotes at page bottom

\usepackage{hyperref}
\usepackage{amsmath}
\usepackage{cleveref}
\usepackage{glossaries}
\usepackage{lipsum}

\hypersetup{
    pdfauthor={Sylvain Robert},
    pdftitle={Local Ensemble Kalman Particle Filters for efficient data  assimilation},
    colorlinks=true,
    citecolor=black,%{blue!10!black},
    linkcolor=black,%
    linkbordercolor=white,
    citebordercolor=white
}

\newacronym{pf}{\textsc{pf}}{Particle Filter}
\newacronym{kf}{\textsc{kf}}{Kalman Filter}
\newacronym{enkf}  {\textsc{e}\textnormal{n}\textsc{kf}}{Ensemble Kalman Filter}
\newacronym{enkpf} {\textsc{e}\textnormal{n}\textsc{kpf}}{Ensemble Kalman Particle Filter}

\newacronym{lenkf}  {\textsc{le}\textnormal{n}\textsc{kf}}{Local \textsc{EnKF}}
\newacronym{block}{\textsc{block-le}\textnormal{n}\textsc{kpf}}{block-local \textsc{e}\textnormal{n}\textsc{kpf}}
\newacronym{naive} {\textsc{naive-le}\textnormal{n}\textsc{kpf}} {naive-local \textsc{e}\textnormal{n}\textsc{kpf}}
\newacronym{global}{\textsc{global-le}\textnormal{n}\textsc{kpf}} {global-local \textsc{e}\textnormal{n}\textsc{kpf}}
\newacronym{lpf}{\textsc{lpf}} {local \textsc{pf}}

\newacronym{ess}{\textsc{ess}}{Effective Sample Size}
\newacronym{mse}{\textsc{mse}}{Mean Squared Error}
\newacronym{msj}{\textsc{msj}}{Mean Squared Jump Size}

\newcommand{\msex}{\textsc{mse}($x$)}
\newcommand{\msedx}{\textsc{mse}($\Delta x$)}
\makeindex             % used for the subject index
\begin{document}

\title*{Localization in High-Dimensional Monte Carlo Filtering}
%\titlerunning{Localization in high-dimensional particle filtering}
% Use \titlerunning{Short Title} for an abbreviated version of
% your contribution title if the original one is too long
\author{Sylvain Robert and Hans R.\ K\"{u}nsch}
% Use \authorrunning{Short Title} for an abbreviated version of
% your contribution title if the original one is too long
%\institute{S.\ Robert \and H.R.\ K\"{u}nsch \at ETH Z\"{u}rich, Seminar for Statistics, Z\"{u}rich, Switzerland, \email{robert@stat.math.ethz.ch}}

\maketitle

%\vspace{-3cm}

\abstract{
    The high dimensionality and computational constraints associated with filtering problems in large-scale geophysical applications are particularly challenging for the \gls{pf}.
    Approximate but efficient methods such as the \gls{enkf} are therefore usually preferred.
    A key element of these approximate methods is localization, which is a general technique to avoid the curse of dimensionality and consists in limiting the influence of observations to neighboring sites.
    However, while it works effectively with the \gls{enkf}, localization introduces harmful discontinuities in the estimated physical fields when applied blindly to the \gls{pf}.
    In the present paper, we explore two possible local algorithms based on the \gls{enkpf}, a hybrid method combining the \gls{enkf} and the \gls{pf}.
    A simulation study in a conjugate normal setup allows to highlight the trade-offs involved when applying localization to \gls{pf} algorithms in the high-dimensional setting. 
    Experiments with the Lorenz96 model demonstrate the ability of the local \gls{enkpf} algorithms to perform well even with a small number of particles compared to the problem size. 
}

\section{Introduction} \vspace*{-10pt}
\label{sec:intro}
\glsresetall 
Monte Carlo methods are becoming increasingly popular for filtering in large-scale geophysical applications, such as reservoir modeling and numerical weather prediction, where they are often called ensemble methods for data assimilation.
The challenging (and interesting) peculiarity of this type of applications is that the state space is extremely high dimensional 
(the number of dimensions of the state $x$ is typically of the order of $10^8$ and the dimension of the observation $y$ of the order of $10^6$),
%(the number of observations is of the order of $10^6$ while the number of dimensions of the state $x$ is of the order of $10^8$),
%(both observation $y$ and state $x$ are of the order of $10^8$),
while the computational cost of the time integration step limits the sample size to less than a hundred. 
Because of those particularly severe constraints, the emphasis is on developing approximate but highly efficient methods, typically relying on strong assumptions and exploiting parallel architectures.

The \gls{pf} provides a fully general Bayesian solution to filtering \cite{gordon_novel_1993,pitt_filtering_1999,doucet_smc_2001}, but it is well-known that it suffers from sample degeneracy and cannot be applied to high-dimensional settings \cite{snyder_obstacles_2008}. 
The most popular alternative to the \gls{pf} in large-scale applications is the \gls{enkf} \cite{evensen_sequential_1994,evensen_ensemble_2003}, a successful but heuristic method, which implicitly assumes that the predictive distribution is Gaussian.

Three main routes for adapting the \gls{pf} to high-dimensional settings can be identified.
The first one is to use an adaptive \gls{pf} with a carefully chosen proposal distribution \cite{pitt_filtering_1999,van_leeuwen_nonlinear_2010}. 
A second approach is to build hybrid methods between the \gls{enkf} and the \gls{pf}, as for example  the \gls{enkpf} \cite{frei_enkpf_2013}.
A third route is localization, as it is a key element of the success of the \gls{enkf} in practice and could avoid the curse of dimensionality \cite{snyder_obstacles_2008,rebeschini_can_2015}.

The first approach requires an explicit model for the transition probabilities, which is typically not available in practical applications. 
Furthermore \cite{snyder_performance_2015} showed that even with the optimal proposal distribution the \gls{pf} suffers from the curse of dimensionality.
Therefore in the present paper we focus on the second and third approaches and explore some possible localized algorithms based on the \gls{pf} and the \gls{enkpf}.
In a simulation study, we extend an example of \cite{snyder_obstacles_2008} to illustrate how localization seemingly overcomes the curse of dimensionality, but at the same time introduces some harmful discontinuities in the estimated state. 
In a second experiment we show how local algorithms can be applied effectively to a filtering problem with the Lorenz96 model \cite{lorenz_optimal_1998}.
The results from these numerical experiments highlight key differences between the algorithms and demonstrate that local \glspl{enkpf} are promising candidates for large-scale filtering applications.

\section{Ensemble filtering algorithms} \vspace*{-10pt}
\label{sec:alg}
Consider a state space model with state process $(x_t)$ and observations $(y_t)$, where the state process evolves according to some deterministic or stochastic dynamics and the observations are assumed to be independent given the state process, with likelihood $p(x_t|y_t)$.
The goal is to estimate the conditional distribution of $x_t$ given $y_{1:t}= (y_1, \dots, y_t)$, called the \emph{filtering} distribution and which we denote by $\pi^f_t$.
In general it is possible to solve this problem recursively by alternating between a \emph{prediction} step where the filtering distribution at time $(t-1)$ is propagated into the predictive distribution $\pi^p_t$ at time $t$, and an \emph{update} step, also called assimilation, where the predictive distribution is updated with the current observation to compute $\pi^f_t$.
The update step is done by applying Bayes' rule as $\pi^f_t(x) \propto \pi^p_t(x) \cdot p(x|y_t)$, where the predictive distribution is the prior and the filtering distribution the posterior to be estimated.

%Sequential Monte Carlo methods 
Sequential Monte Carlo methods \cite{doucet_smc_2001} 
approximate the predictive and filtering distributions by  finite samples, or ensembles of \emph{particles}, denoted by $(x^{p,i}_t)$ and $(x^{f,i}_t)$ respectively, for $i=1,\dots, k$.
The update step consists in transforming the predictive ensemble $(x^{p,i}_t)$ into an approximate sample from the filtering distribution $\pi^f_t$.
We briefly present the \gls{pf} and \gls{enkf} in this context and give an overview of the \gls{enkpf}.
Henceforth we consider the update step only and drop the time index $t$. Additionally for the \gls{enkf} and \gls{enkpf} we assume that the observations are linear and Gaussian, i.e.\ $p(x|y) = \phi (y;\ Hx, R)$, 
the Gaussian density with mean $Hx$ and covariance $R$ evaluated at $y$. 
%with $\phi(y;\ a,b)$ the Gaussian density with mean $a$ and covariance $b$ evaluated at $y$. 

%--------------
\runinhead{The \acrlong{pf}} approximates the filtering distribution as a mixture of point masses at the predictive particles, reweighed by their likelihood. More precisely:
\begin{equation}
    \hat{\pi}^f_{PF} (x) = \sum_{i=1}^k w_i \, \delta_{x^{p,i}} (x), 
    \qquad w_i \propto \phi(y;\ Hx^{p,i}, R).
\end{equation}
A non-weighted sample from this distribution can be obtained by resampling, for example with a balanced sampling scheme \cite{kunsch_recursive_2005}.
The \gls{pf} is asymptotically correct  (also for non-Gaussian likelihoods), but to avoid sample degeneracy it needs a sample size which increases exponentially with the size of the problem (for more detail see \cite{snyder_obstacles_2008}).

%--------------
\runinhead{The \acrlong{enkf}} is a heuristic method which applies a \gls{kf} update to each particle with stochastically perturbed observations.
% (\emph{square-root} schemes exist but are not considered here).
More precisely it constructs $(x^{f,i})$ as a balanced sample from the following Gaussian mixture:

\begin{equation}
    \hat{\pi}^f_{EnKF} (x) = \sum_{i=1}^k \frac{1}{k} \, \phi(x;\ x^{p,i} + \hat{K} (y - Hx^{p,i}), \hat{K}R\hat{K}'),
\end{equation}
where $\hat{K}$ is the Kalman gain estimated with $\hat{\Sigma}^p$, the sample covariance of $(x^{p,i})$:
%the sample covariance $\hat{\Sigma}^p$: 
% uncut:
The stochastic perturbations of the observations are added to ensure that the filter ensemble has the correct posterior covariance on expectation. 
%Some variants of the algorithm use a square-root scheme such that the filter ensemble has the exact correct posterior covariance, but such methods are out of the scope of the present paper (see for example
%%\cite{hunt_efficient_2007}).
% \cite{whitaker_ensemble_2002,tippett_ensemble_2003,hunt_efficient_2007}).
%% \cite{whitaker_ensemble_2002}).

%--------------
\runinhead{The \acrlong{enkpf}} combines the \gls{enkf} and the \gls{pf} by decomposing the update step into two stages as $\pi^f(x) \propto \pi^p(x) \cdot p(x|y)^{\gamma} \cdot p(x|y)^{1-\gamma}$, following the progressive correction idea of \cite{musso2001improving}. The first stage, going from $\pi^p(x)$ to $\pi^{\gamma}(x) \propto \pi^p(x) \cdot p(x|y)^{\gamma}$  is done with an \gls{enkf}. The second stage is done with a \gls{pf} and goes from $\pi^{\gamma}(x)$ to $\pi^f(x) \propto \pi^{\gamma}(x) \cdot p(x|y)^{1-\gamma}$.  %followed by a \gls{pf} correction. 
The resulting posterior distribution can be derived analytically as the following weighted Gaussian mixture:
\begin{equation}
    \hat{\pi}^f_{EnKPF} (x) = \sum_{i=1}^k \alpha^{\gamma,i} \, \phi( x;\ \mu^{\gamma,i}, \Sigma^{\gamma}),
\end{equation}
where the expressions for the parameters of this distribution and more details on the algorithm can be found in \cite{frei_enkpf_2013}.
To produce the filtering ensemble $(x^{f,i})$, one first samples the mixture components with probability proportional to the weights $\alpha^{\gamma,i}$, using for example a balanced sampling scheme, and then adds an individual noise term with covariance $\Sigma^{\gamma}$ to each particle.
The parameter $\gamma$ defines a continuous interpolation between the \gls{pf} ($\gamma=0$) and the \gls{enkf} ($\gamma=1$). 
In the present study the value of $\gamma$ is either fixed, for the sake of comparison, or chosen adaptively. In the later case $\gamma$ is chosen such that the equivalent sample size of the filtering ensemble is within some acceptable range. 
Alternative schemes for choosing $\gamma$ such as minimizing an objective cost function are currently being investigated but are beyond the scope of this work.

\section{Local algorithms}  \vspace*{-10pt}
Localization consists essentially in updating the state vector by ignoring long range dependencies. 
This is a sensible thing to do in geophysical applications where the state represents discretized spatially correlated fields of physical quantities.
By localizing the update step and using  local observations only, one introduces a bias, but achieves a considerable gain in terms of variance reduction for finite sample sizes.
For local algorithms the error is asymptotically bigger than for a global algorithm, but it is not dependent on the system dimension anymore and therefore avoids the curse of dimensionality.
Furthermore, local algorithms can be efficiently implemented in parallel  and thus take advantage of modern computing architectures. 

The \gls{lenkf} consists in applying a separate \gls{enkf} at each site, but limiting the influence of the observations to sites that are spatially close (there are different ways to accomplish this in practice, see for example \cite{houtekamer_sequential_2001,ott_lenkf_2004,hunt_efficient_2007}). 
Analogously, we define the \gls{lpf} as a localized version of the \gls{pf}, where the update is done at each location independently, considering only observations in a ball of radius $\ell$. 
In order to avoid arbitrary ``scrambling'' of the particles indices, we use a balanced sampling scheme \cite{kunsch_recursive_2005}, and some basic ad-hoc methods to reduce the number of discontinuities, but we do not solve this problem optimally as it would greatly hinder the efficiency of the algorithm.

For the \gls{enkpf} we define two different local algorithms: the \gls{naive}, in which localization is done exactly as for the \gls{lenkf}, and the \gls{block}, in which the observations are assimilated sequentially but their influence is restricted to a local area. The \gls{naive} does not take particular care of the introduced discontinuities beyond what is done for the \gls{pf}, but it is straightforward to implement. 
The \gls{block}, on the other hand, uses conditional resampling in a transition area surrounding the local assimilation window, which ensures that there are no sharp discontinuities, but it involves more overhead computation. For more detail about the local \gls{enkpf} algorithms see \cite{robert_arxiv_2016}.

\section{Simulation studies} \vspace*{-10pt}
We conducted two simulation studies: first a one-step conjugate normal setup where the effect of localization can be closely studied, and second a cycled experiment with the Lorenz96 model, a non-linear dynamical system displaying interesting non-Gaussian features. 

\subsection{Conjugate normal setup} \vspace*{-10pt}
\label{sec:sim}
We consider a simple setup similar to the one in \cite{snyder_obstacles_2008}, with a predictive distribution $\pi^p$ assumed to be a $N$-dimensional normal with mean zero and covariance $\Sigma^p$.
To imitate the kind of smooth fields that we encounter in geophysical applications, we construct the covariance matrix as $\Sigma^p_{ii}=1$ and $\Sigma^p_{ij}=K_{GC}( d(i,j)/r)$, where $K_{GC}$ is the Gaspari-Cohn kernel \cite{gaspari_construction_1999}, $d(i,j)$ the distance between sites $i$ and $j$ on a one-dimensional domain with periodic boundary conditions, and the radius $r$ in the denominator is chosen such that the covariance has a finite support of 20 grid points.
From this process we generate observations of every component of $x$ and standard Gaussian noise:
\begin{align}
    x       \sim \mathcal{N} (0, \Sigma^p), \label{eq:sim_pred} \qquad 
    y | x   \sim \mathcal{N} (x, I).
\end{align}

% uncut:
%An example of a realization of this process can be seen as the dark blue line in \cref{fig:ens_ex} with the observations as red crosses.

In order to study the finite sample properties of the different algorithms, we compute the \gls{mse} of the ensemble mean in estimating the value $x$ at each location, which we denote by \msex{}. Because the prior is conjugate to the likelihood, we can  compute the \msex{} of the posterior mean analytically for known $\Sigma^p$ as the trace of the posterior covariance matrix and use this as a reference.
For the simulation we use a sample size of $k=100$ and average the results over 1000 runs.
It should be noted that because the predictive distribution is normal, this setup is favorable to the \gls{enkf} and \gls{lenkf}, but the \glspl{enkpf} should still perform adequately. 
For the local algorithms the localization radius $\ell$ was set to 5, resulting in a local window of 11 grid points, which is smaller than the correlation length used to generate the data. Later on we study the effect of $\ell$ on the performance of the algorithms. For the \gls{enkpf} algorithms the parameter $\gamma$ was fixed to 0.25, which means a quarter of \gls{enkf} and three-quarter of \gls{pf}. In practice one would rather choose the value of $\gamma$ adaptively, but the exact value does not influence the qualitative conclusions drawn from the experiments and fixing it in this way makes the comparison easier.

\begin{figure}
    \centering
    \includegraphics[width=.75 \textwidth]{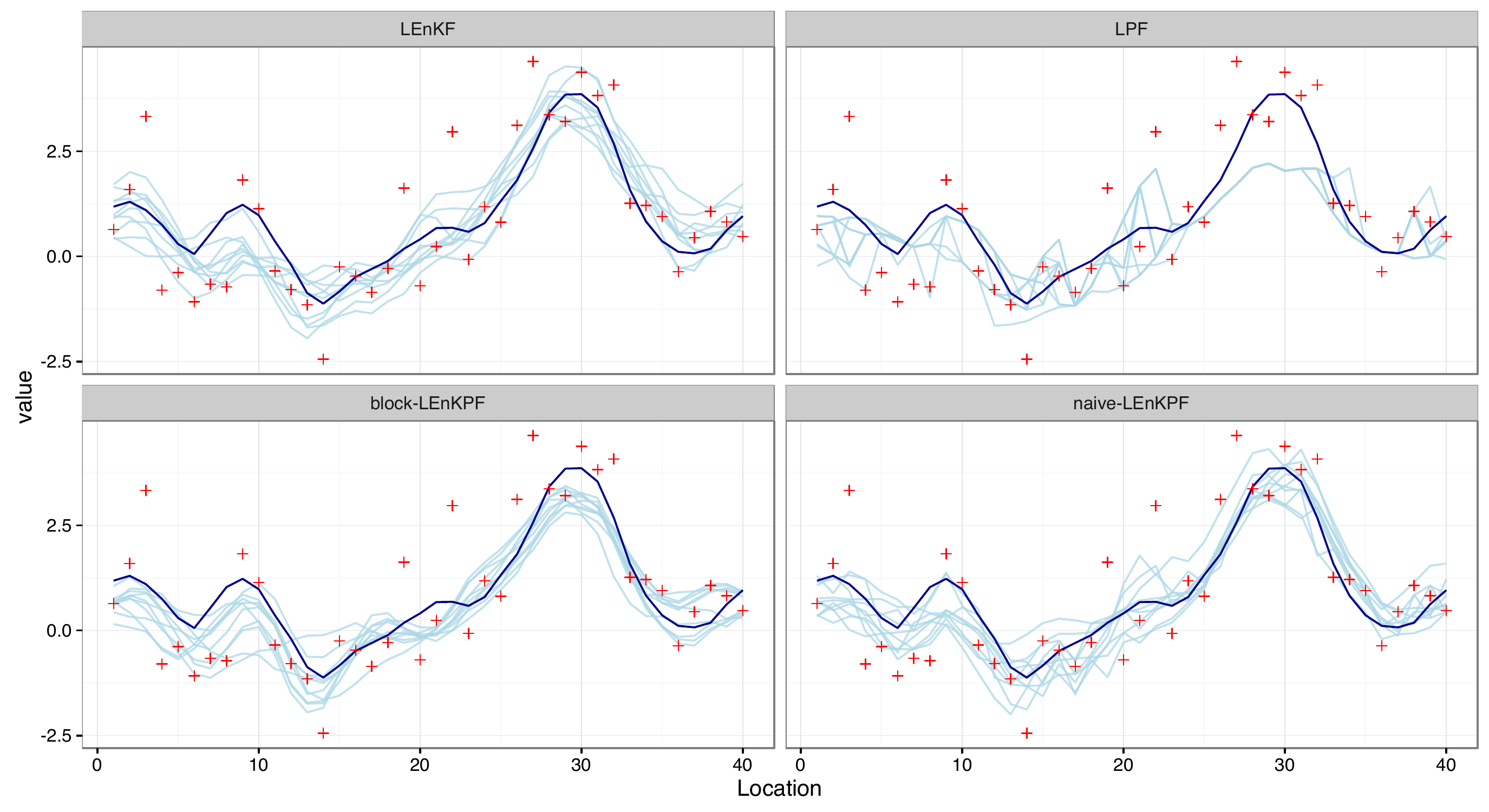}
    \caption{Example of analysis ensemble with different local algorithms. Each particle is a light blue line, the true state in dark and the observations in red. The ensemble size is restricted to 10 and the domain size to 40 for better legibility.}
    \label{fig:ens_ex}
\end{figure}

% uncut
An example of a sample from the filtering distribution produced by different local algorithms is shown in \cref{fig:ens_ex}, with each particle represented as a light blue line, the true state in dark and the observations in red. For more clarity the ensemble size is set to 10 and the system dimension to 40. While all algorithms manage to recover more or less the underlying state, it is clear that they vary in terms of quality. The \gls{lpf} in particular suffers from sample depletion, even when applied locally, and displays strong discontinuities. If one looks closely at the \gls{naive} ensemble, discontinuities can also be identified. The \gls{block} and the \gls{lenkf}, on the other hand, produce smoother posterior particles. This example is useful to highlight the behavior of the different local algorithms qualitatively, but we now proceed to a more quantitative assessment with a repeated simulations experiment.

\begin{figure}
    \centering
    \includegraphics[width=.75 \textwidth]{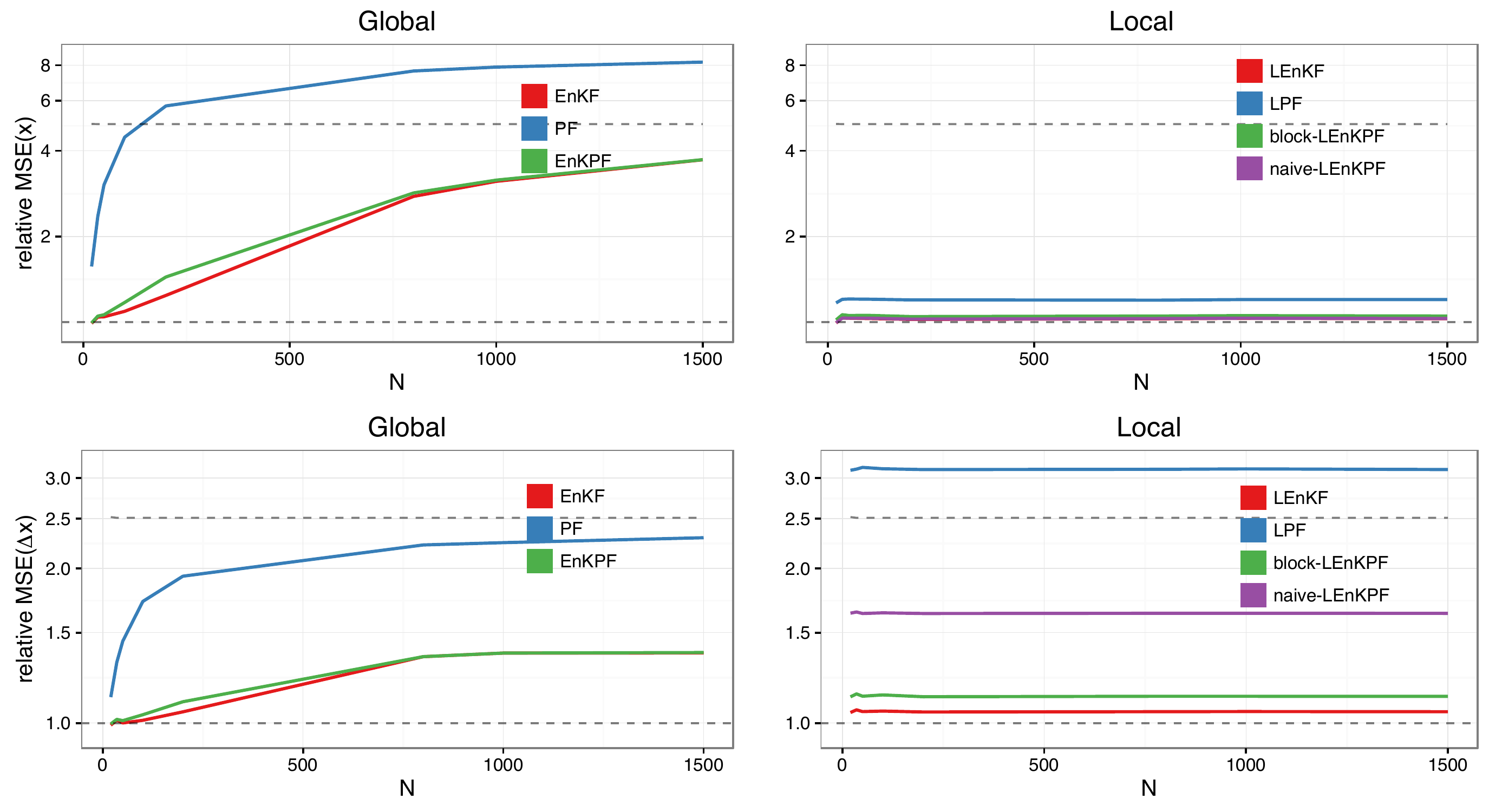}
    \caption{Illustration of the relationship between the system dimension $N$ and different quantities. In the first row is the \msex{} for the global algorithms on the left and the local algorithms on the right. In the second row the same but for the \msedx{}.
    All the values are given relative to the optimal one obtained with the true posterior distribution (dashed line at 1). The relative \gls{mse} of using the prior without using any observation is given by the second dashed line. Notice the log-scale on the y-axis.}
    \label{fig:mse}
\end{figure}

\runinhead{Beating the curse of dimensionality:} 
In the first row of \cref{fig:mse}, the \msex{} is plotted as a function of the system dimension $N$, for the global algorithms on the left and the local algorithms on the right.
%In the first row of \cref{fig:mse} the \msex{} for the global algorithms on the left and the local algorithms on the right is plotted as a function of the system dimension $N$, averaged over the 1000 simulation runs. 
The values are normalized by the optimal \msex{} to make them more interpretable.
The \gls{pf} degenerates rapidly, with an \msex{} worse than using the prior mean (upper dashed line). 
The \gls{enkf} and the \gls{enkpf} suffer as well from the curse of dimensionality, although to a lesser extent. 
%It is clear that the \gls{pf} degenerates rapidly, while the \gls{enkf} and the \gls{enkpf} suffer as well from the curse of dimensionality, although to a lesser extent. 
The local algorithms, on the other hand, are immune to the increase of dimensions $N$ and their \msex{} is constant and very close the optimum, which confirms that localization is working as expected. The \gls{lenkf}, \gls{naive} and \gls{block} make an error of less than 5\% while the \gls{lpf} is 20\% worse than the optimum.

\runinhead{The cost of localization:}
As the old statistical adage goes, there is no free lunch: localization comes at a cost, particularly for \gls{pf} algorithms.
When doing the update locally with the \gls{enkf}, the filtering samples are relatively smooth fields, because the update applies spatially smoothly varying corrections to the predictive ensemble.
However, for the \gls{lpf}, when different particles are resampled at neighboring sites, arbitrarily large discontinuities can be created.
While this might be discarded as harmless, it is not the case when the fields of interest are spatial fields of physical quantities used in numerical solvers of partial differential equations. 
One way to measure the impact of discontinuities is to look at the \gls{mse} in estimating the lag one increments $\Delta x$, which we denote as \msedx{}. While the \msex{} is computed for the posterior mean, the \msedx{} is computed for each particle separately and then averaged. We again compute the expected \msedx{} under the conjugate posterior distribution and use it as reference.

%the first spatial derivative, which we denote as \msedx{}. Given a particular realization of the process one can compute the optimal \msedx{} analytically and use it as a reference.

The plots in the second row of \cref{fig:mse} show this quantity for the different algorithms averaged over 1000 simulation runs. 
The \msedx{} of the local algorithms is still constant as a function of $N$, as expected, but in the cases of the \gls{naive} and the \gls{lpf} its value is worse than for the respective global algorithms. %, by 20\% and 300\% respectively. This large \msedx{} comes from the added discontinuities due to local resampling. 
On the other hand, the \gls{lenkf} and the \gls{block} improve on their global counterparts and have an error relatively close to the optimum.

\begin{figure}
    \centering
    \includegraphics[width=.75 \textwidth]{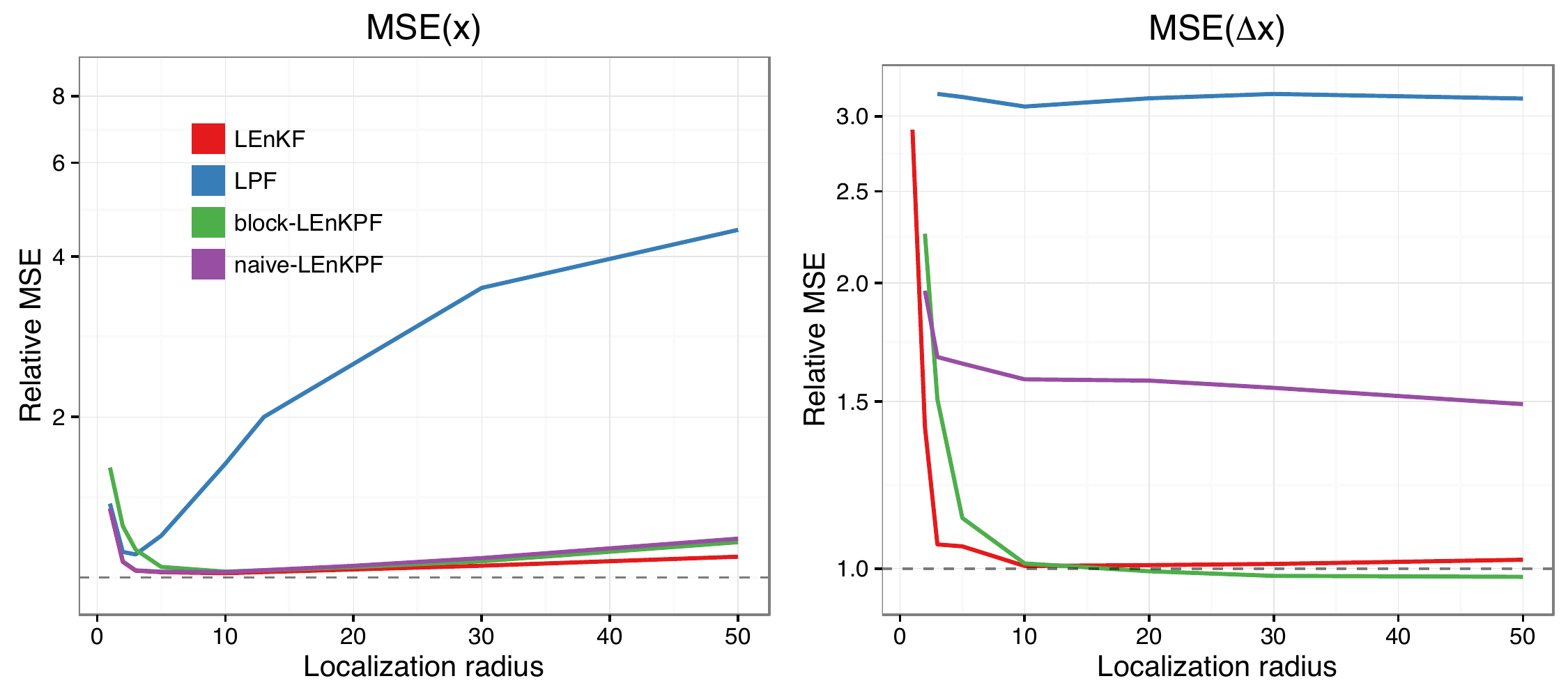}
    \caption{Trade-off of localization: influence of $\ell$ on \msex{} and \msedx{} for the local algorithms. The ensemble size $k$ was fixed to 100 and the system dimension $N$ to 200. Notice the log-scale on the y-axis.}
    \label{fig:mse_kl}
\end{figure}

\runinhead{Localization trade-off:}
In the previous experiment we fixed $\ell$, the localization radius, to 5, and looked at what happens in terms of prediction accuracy with the \msex{}, and in terms of discontinuities with the \msedx{}. 
%In \cref{fig:mse_kl}, we  fix $N$ to 200, keep $k=100$ and 
%and look at results as a function of $\ell$. 
In \cref{fig:mse_kl} we now look at \gls{mse} as a function of $\ell$, fixing $N$ to 200 and $k$ to 100.
For large values of $\ell$ the \msedx{} is smallest as discontinuities are avoided, but the \msex{} is not optimal, particularly for the \gls{lpf}. As $\ell$ is reduced the \msex{} decreases for all methods, while \msedx{} is kept constant for a wide range of $\ell$ values. At some point, different for each algorithm, the localization is too strong and becomes detrimental, with both \msex{} and \msedx{} sharply increasing.
%In the left panel of \cref{fig:mse_kl} one can see that the \msex{} initially decreases for all algorithms, before increasing slowly for most algorithms, or very sharply for the \gls{lpf}, which even goes off-chart. 
%The evolution of the \msedx{} on the right panel shows an initial sharp decrease followed by a plateau. 
This behavior illustrates the trade-off at hand when choosing the localization radius:
picking a too small value introduces a bias by neglecting useful information and creates too much discontinuities, while choosing a too large value does not improve \msedx{} but leads to poorer performance in terms of \msex{}.

%% uncut:
%\todo{add the uncunt chapter in personal version}
%\input{BAYSM16_supplement}

\subsection{Filtering with the Lorenz96 model} \vspace*{-10pt}
The Lorenz96 model \cite{lorenz_optimal_1998} is a 40-dimensional non-linear dynamical system which displays a rich behavior and is often used as a benchmark for filtering algorithms. In \cite{frei_enkpf_2013} it was shown that the \gls{enkpf} outperforms the \gls{lenkf} in some setups of the Lorenz96 model, but the sample size required was of 400. In the present experiment we use the same setup as in \cite{frei_enkpf_2013} but with much smaller and realistic ensemble sizes. The data are assimilated at time intervals of 0.4, which leads to strong non-linearities and thus highlights better the relative advantages of the \gls{enkpf}. Each experiment is run for 1000 cycles and repeated 20 times, which provides us with stable estimates of the average performance of each algorithm. As in \cite{frei_enkpf_2013}, the parameter $\gamma$ of the \glspl{enkpf} is chosen adaptively such that the equivalent sample size is between 25 and 50\% of the ensemble size. It should be noted that for local algorithms, a different $\gamma$ is chosen at each location, which provides added flexibility and allows to adapt to locally non-Gaussian features of the distribution.
%We consider \msex{} only and do not look at \msedx{} because the system being integrated in time the quality of the estimation of the derivative is already taken into account by looking at \msex{}, which we refer to as \gls{mse}.
We consider \msex{} only and denote it simply by \gls{mse}. It also takes  errors in the estimation of increments into account through integration in time during the propagation steps.

\begin{figure}
    \centering
    \includegraphics[width= .75 \textwidth]{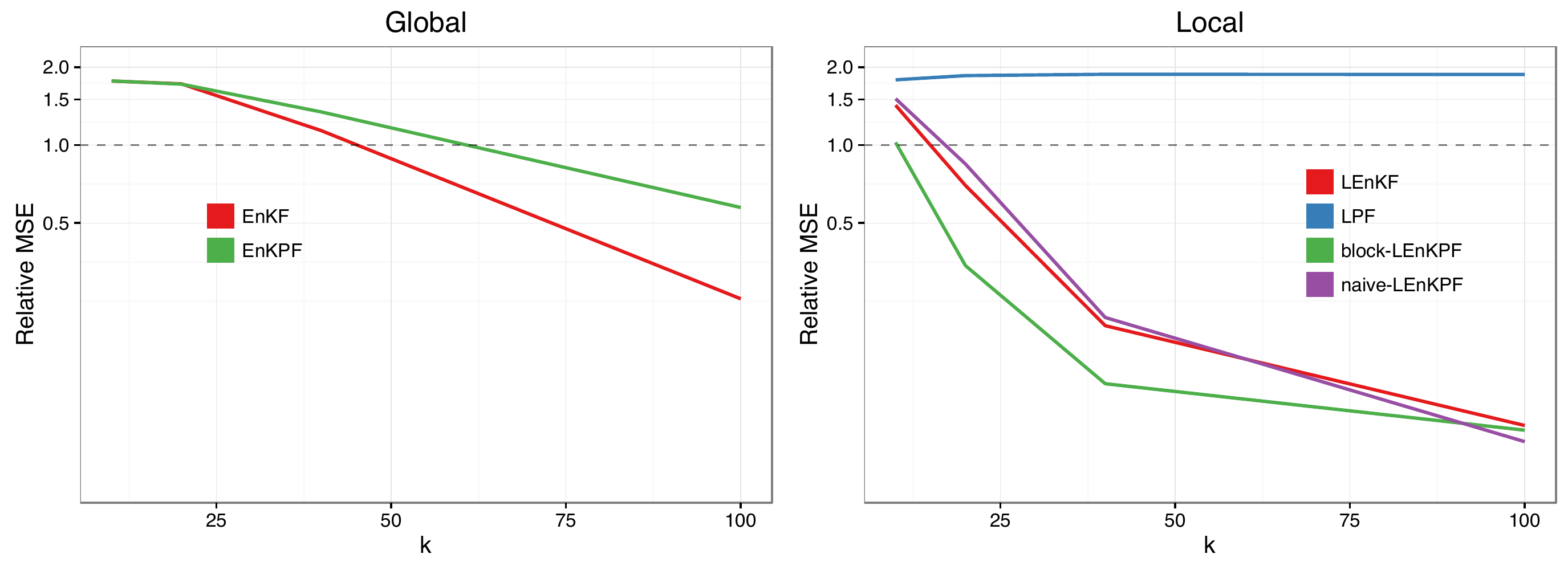}
    \caption{Global and local algorithms results with the Lorenz96 model. On the y-axis the \gls{mse} and on the x-axis increasing ensemble sizes $k$. There is no global \gls{pf} as the algorithm does not work for such small ensemble sizes. The \gls{mse} is computed relative to the \gls{mse} of an ensemble of the same size but which does not assimilate any observation. Notice the log-scale on the y-axis.}
    \label{fig:lorenz_local}
\end{figure}

In the left panel of \cref{fig:lorenz_local} the \gls{mse} of the global algorithms is plotted against ensemble size. The \gls{pf} is not  represented as it diverges for such small values. The \gls{mse} is computed relative to the performance of the prior, which is simply taken as the \gls{mse} of an ensemble of the same size evolving according to the dynamical system equations but not assimilating any observations. 
With ensemble sizes smaller than 50, the filtering algorithms are not able to do better than the prior, which means that trying to use the observations actually makes them worse than not using them at all. Only for ensemble sizes of 100 and more do the global algorithms start to become effective. In practice we are interested in situations where the ensemble size is smaller than the system dimension (here 40), and thus the global methods are clearly not applicable. 

On the right panel of \cref{fig:lorenz_local} we show the same plot but for the local algorithms. For sample sizes as small as 20 or 30 the performances are already quite good. The \gls{lpf}, however, does not work at all, probably because it still suffers from sample depletion and because the discontinuities it introduces have a detrimental impact during the prediction step of the algorithm. The \gls{block} clearly outperforms the other algorithms, particularly for smaller sample sizes. This indicates that it can localize efficiently the update without harming the prediction step by introducing discontinuities in the fields.

\begin{figure}
    \centering
    \includegraphics[width= .75 \textwidth]{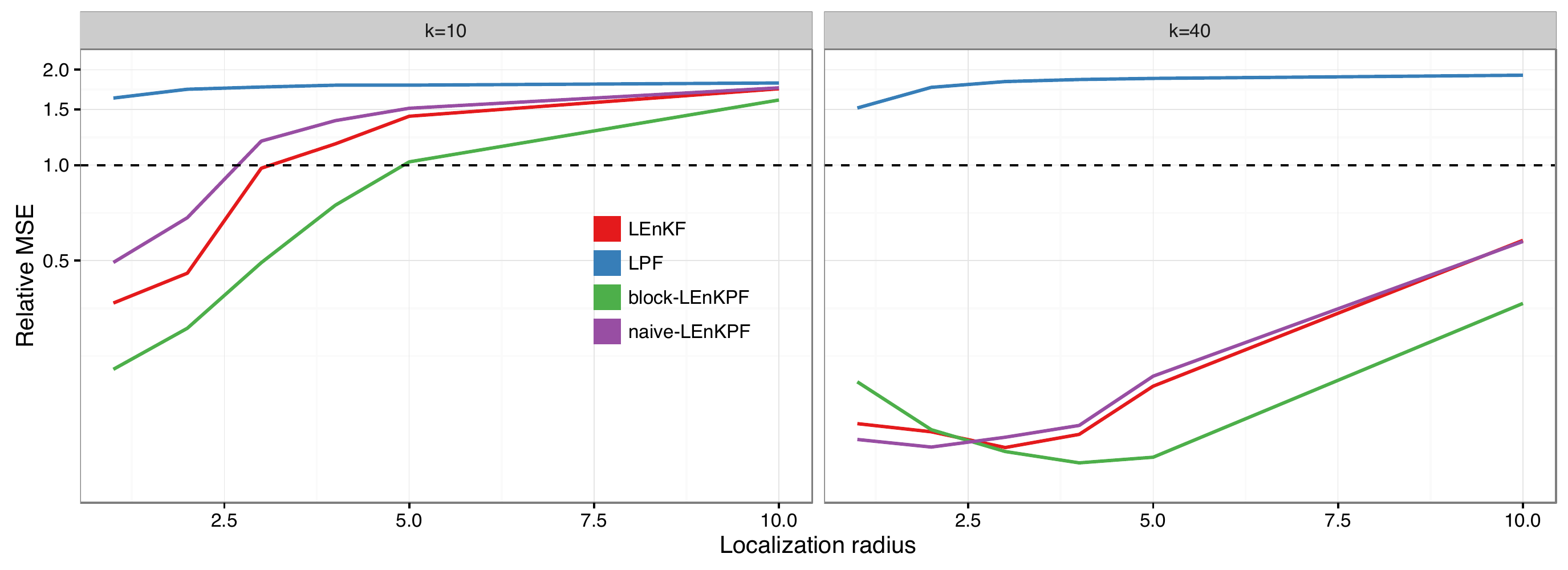}
    \caption{Interplay of ensemble size $k$ and localization radius $\ell$ for the the Lorenz96 model. The relative \gls{mse} is plotted as a function of $\ell$ for different value of $k$ in the different panels. Notice the log-scale on the y-axis.}
%        Each subplot is for a different $k$ value, while the x-axis is for increasing $\ell$ and the y-axis the relative \gls{mse}.}
    \label{fig:lorenz_kl}
\end{figure}

In order to better highlight the trade-off of localization, we plot similar curves but as a function of the localization radius $\ell$ in \cref{fig:lorenz_kl}. One can see that for small $k$ (left panel), the error is increasing with $\ell$, which shows that localization is absolutely necessary for the algorithm to work. For $k=40$ (right panel), the \gls{mse} first decreases and then increases, with an optimal $\ell$. 
Experiments with larger values display curves that get flatter and flatter as $k$ increases, showing that as the ensemble size is larger, the localization strength needed is smaller, as expected.

\section{Conclusion} \vspace*{-10pt}
% The return of the Jedi
\label{sec:conclusion}
Localization is an effective tool to address some of the difficulties associated with high-dimensional filtering in large-scale geophysical applications.
Methods such as the \gls{enkf} can be localized easily and successfully as they vary smoothly in space.
At first sight, the \gls{lpf} does seem to overcome the curse of dimensionality; however, looking more carefully, one notices that it introduces  harmful discontinuities in the updated fields.
The two localized \glspl{enkpf} both overcome the curse of dimensionality and handle better the problem of discontinuities.

The simple conjugate example studied in this paper highlighted the potential improvements coming from localization, as well as the pitfalls when applied blindly to the \gls{pf}. 
The trade-off between the bias coming from localization and the gain coming from the reduced variance was illustrated by exploring the behavior of the algorithms as a function of the localization radius $\ell$. 
Experiments with the Lorenz96 model showed that local algorithms can be successfully applied with ensemble sizes as small as 20 or 30, and highlighted  the localization trade-off.
In particular, the \gls{block} fared remarkably well, outperforming both the \gls{naive} and the \gls{lenkf} in this challenging setup.
This confirms other results that we obtained with more complex dynamical models mimicking cumulus convection \cite{robert_arxiv_2016} and encourages us to pursue further research with localized \glspl{enkpf} in a large-scale application in collaboration with Meteoswiss.

\bibliography{baysm_bib}{} 
\bibliographystyle{spmpsci}
\end{document}